\documentclass[a4paper,aps,prl,twocolumn,showpacs,preprintnumbers,amsmath,amssymb]{revtex4}
\usepackage{bbm}
\usepackage{amsmath}
\usepackage{verbatim}
\usepackage{graphicx}
\newcommand{\tr}[1]{\mathrm{tr}\{#1\}}
\newcommand{\ud}{\mathrm{d}}

\begin{document}

\title{Quantum benchmark for storage and transmission of coherent states}

\author{K. Hammerer$^1$, M.M. Wolf$^1$, E.S. Polzik$^{2,3}$, J.I. Cirac$^1$}

\affiliation{$^1$Max-Planck--Institut f\"ur Quantenoptik, D-85748 Garching, Germany\\
$^2$QUANTOP, Danish Research Foundation Center for Quantum Optics\\
$^3$Niels Bohr Institute, DK 2100 Copenhagen, Denmark}

\begin{abstract}
We consider the storage and transmission of a Gaussian distributed
set of coherent states of continuous variable systems. We prove a
limit on the average fidelity achievable when the states are
transmitted or stored by a classical channel, i.e., a measure and
repreparation scheme which sends or stores classical information
only. The obtained bound is tight and serves as a benchmark which
has to be surpassed by quantum channels in order to outperform any
classical strategy. The success in experimental demonstrations of
quantum memories as well as quantum teleportation has to be judged
on this footing.
\end{abstract}


\maketitle

Coherent distribution, storage and manipulation of quantum states
is a technical challenge which received extensive theoretical and
experimental interest in the last years stimulated by the promises
of quantum information science \cite{NC}. A wide class of schemes
can be very generally understood as an attempt to establish a
channel for the reliable transmission of quantum states. This
applies in particular to quantum teleportation \cite{BBC, VBK,
ATOMS, PHOTONS, FBFKP, ZGCLK, BTBSRBSL, YAF}, where states are
sent through an entanglement assisted classical channel, but just
as well to the concept of a quantum memory \cite{KMP, SJSP, JSCFP,
FL, LDBH, BZL, DP, CZKM}, where the channel acts in time rather
than in space and the accent is on the state transfer between
light and atoms. Concerning the reliable transmission and storage
of quantum states it is clear that in the ideal case a quantum
channel will always surpass a classical channel, i.e. any strategy
where the quantum state is measured, the corresponding classical
data stored or transmitted and then used to reconstruct the
initial state as good as possible \cite{fn}. However, under
realistic conditions a quantum channel suffers inevitably from
imperfections such that it might become possible to achieve the
same effect by means of a classical channel. Therefore there is
need for a criterion which allows to distinguish an imperfect
quantum channel from a perfect classical channel and justifies
proclamation of success in the experimental demonstration of
quantum teleportation and quantum memories.

Such a criterion has been derived some time ago for channels
acting on finite dimensional systems \cite{P, MP, DBE, BEM} and
found applications in seminal experiments on quantum teleportation
with single photons \cite{PHOTONS} and ions \cite{ATOMS}. For
channels acting on infinite dimensional systems a corresponding
criterion was \textit{conjectured} some time ago in \cite{FBFKP,
BFK}. Though a proof for this criterion was yet to be found the
claim for successful teleportation was based exclusively on this
ground in several, likewise seminal teleportation experiments
using EPR-squeezed light \cite{FBFKP, ZGCLK, BTBSRBSL, YAF}. The
same criterion was applied very recently to a quantum memory
experiment \cite{JSCFP} where coherent states of light were stored
in the collective spin of atoms.

In this paper we solve this longstanding problem and provide a
rigorous proof for the criterion conjectured in \cite{FBFKP, BFK}.
This puts the central claims of experiments \cite{FBFKP, ZGCLK,
BTBSRBSL, YAF, JSCFP} - to have demonstrated a quantum gain in the
transmission or storage of coherent states - on logically firm
grounds. We emphasize that only now the success of these key
experiments is rigorously validated. Moreover, the present result
gives a solution to the state estimation problem for coherent
states and leads to a closed expression for the accessible
fidelity introduced in \cite{AFKW}.

The paper is organized as follows: We first characterize a general
classical channel mathematically and state the optimization
problem to be solved here. Then we present this solution and give
an elementary and rigorous proof. We close by relating the result
to other work.

The figure of merit in terms of which the quality of a channel is
quantified is usually taken to be the average fidelity achieved
when the channel acts on a predefined set of input states
\cite{BTBSRSL}. Let $\{|\psi_x\rangle\}$ be this set and let an
input $|\psi_x\rangle$ occur with a probability $p(x)$. If the
channels's output is $E(|\psi_x\rangle)$ then the average fidelity
is defined as
\begin{equation*}
\bar{F}=\sum_x p(x)\langle\psi_x|E(|\psi_x\rangle)|\psi_x\rangle.
\end{equation*}
This number is equal to one only for the ideal channel
transmitting or storing every state perfectly. Now the task is to
find the maximal value of $\bar{F}$ achievable with a classical
channel, i.e. to identify the optimal measure-and-prepare
strategy. Any channel yielding a higher average fidelity is then
necessarily quantum in the sense that it outperforms every
strategy which is based on the mere storage or transmission of
classical information. Any classical channel can be described by a
POVM \cite{NC} measurement $\{M_y\}$ where $y$ denotes the outcome
occurring with a probability $|\langle\psi_x|M_y|\psi_x\rangle|^2$
and a reconstruction rule $y\rightarrow\rho_y$ determining which
state $\rho_y$ is prepared when $y$ was the measurement outcome.
The channel then acts as
\begin{equation*}
E(|\psi_x\rangle)=\sum_y|\langle\psi_x|M_y|\psi_x\rangle|^2\rho_y.
\end{equation*}
The fidelity bound for classical channels relative to the set of
input states $\{|\psi_x\rangle\}$ is then
\begin{equation}\label{fmax}
F_\mathrm{max}=\sup_{M_y}\sup_{\rho_y}\sum_x\sum_y
p(x)\langle\psi_x|M_y|\psi_x\rangle|^2\langle\psi_x|\rho_y|\psi_x\rangle.
\end{equation}
This optimization problem is known under the title state
estimation in the theory of quantum detection and has in fact a
long history \cite{YuenLax, Ho, He, YKL}. From the plethora of
results known in this field the one concerning channels acting on
$\mathbb{C}^2$ and an input set consisting of all pure states with
a uniform distribution over the Bloch sphere received particular
practical relevance in the last years. In \cite{MP} it was shown
that for this case $F_\mathrm{max}=2/3$. This value was the
appropriate benchmark in several teleportation experiments using
single photons \cite{PHOTONS} and recently also trapped ions
\cite{ATOMS} and was beaten by measured fidelities ranging from
$.70$ to $.89$ in \cite{PHOTONS} and from $.75$ to $.78$ in
\cite{ATOMS} proving the presence and necessity of entanglement in
these experiments.

Less is known for channels acting on an infinite dimensional
Hilbert space. Despite the increasing importance of coherent
states $\{|\alpha\rangle\}$ for quantum communication and in
particular quantum cryptography by now no classical-quantum bound
has been proven for channels acting on these states. In \cite{BFK}
it was shown that if the coherent states are distributed in phase
space according to a Gaussian distribution
\mbox{$p(\alpha)=\lambda/\pi\exp(-\lambda|\alpha|^2)$}  an average
fidelity \mbox{$\bar{F}=(1+\lambda)/(2+\lambda)$} can be achieved
by means of a heterodyne measurement, described by a POVM
$\{|\alpha\rangle\langle\alpha|/\pi\}$, and the preparation of
appropriate coherent states. It was conjectured there that this
might  be optimal but since then this question remained open. In
fact, in the state estimation problem with minimum mean square
error  the heterodyne measurement turned out to be optimal
\cite{YuenLax}. However, this  problem is different from the
present one with respect to the figure of merit and due to the
fact that in \cite{YuenLax} the reconstruction of the state, which
is crucial in our context, is not considered. Nevertheless, the
value of $1/2$ attained for the flat distribution
($\lambda\rightarrow 0$) was used as a criterion to verify
teleportation in experiments \cite{FBFKP, ZGCLK, BTBSRBSL, YAF}
using EPR-squeezed light where measured average fidelities range
from $.58$ to $.64$.

In the following we will settle this question by proving that for
any classical strategy
\begin{equation}\label{bound}
F_\mathrm{max}\leq\frac{1+\lambda}{2+\lambda}
\end{equation}
holds necessarily. Moreover, this bound is tight since it can be
achieved by means of the strategy derived in \cite{BFK}, and thus
equality holds in (\ref{bound}). This is the main result of this
paper.

The proof we are going to present now is elementary and we start
by simplifying and conveniently reformulating the problem. The
first simplification relies on the fact that without loss of
generality we can restrict the optimization in equation
(\ref{fmax}) to POVMs consisting of projectors
\mbox{$M_y=|\phi_y\rangle\langle\phi_y|$} ($|\phi_y\rangle$ not
necessarily normalized) and also to pure states
$\rho_y=|\chi_y\rangle\langle\chi_y|$. This is easily seen by
noting that we can always decompose the POVM elements
\mbox{$M_y=\sum_v|m_{v,y}\rangle\langle m_{v,y}|$} and similarly
the states \mbox{$\rho_y=\sum_w q_{w,y}|r_{w,y}\rangle\langle
r_{w,y}|$} such that we can write the average fidelity as
\begin{equation*}
\bar{F}=\sum_x\sum_{y,v,w}
p(x)|\langle\psi_x|\sqrt{q_{w,y}}|m_{v,y}\rangle|^2|\langle\psi_x|r_{w,y}\rangle|^2.
\end{equation*}
Absorbing the redundant parameters $v,\,w$ into $y$ and
identifying \mbox{$\sqrt{q_{w,y}}|m_{v,y}\rangle$} and
\mbox{$|r_{w,y}\rangle$} with $|\phi_y\rangle$ and
$|\chi_y\rangle$ respectively we see that for any POVM there
exists always another one which has the desired properties and
yields the same average fidelity.

We therefore have for coherent input states $\{|\alpha\rangle\}$
with a Gaussian distribution
\mbox{$p(\alpha)=\lambda/\pi\exp(-\lambda|\alpha|^2)$}
$$
\bar{F}=\sum_y\int\! d\alpha
p(\alpha)|\langle\alpha|\phi_y\rangle|^2|\langle\alpha|\chi_y\rangle|^2.
$$
Note that the sum over $y$ stands symbolically for sums or
integrations over a suitable measurable set. Using this expression
for $\bar{F}$ and defining
$$
A_{\phi_y}=\int \ud\alpha
p(\alpha)|\langle\alpha|\phi_y\rangle|^2|\alpha\rangle\langle\alpha|.
$$
equation (\ref{fmax}) can be reformulated more compactly as
\begin{equation}\label{fmax2}
F_\mathrm{max}=\sup_{|\phi_y\rangle}\sup_{|\chi_y\rangle}\sum_y\langle\chi_y|A_{\phi_y}|\chi_y\rangle=\sup_{|\phi_y\rangle}\sum_y||A_{\phi_y}||_\infty.
\end{equation}
The optimization with respect to the reconstructed states
$|\chi_y\rangle$ is trivial and implicitly performed in the last
identity by noting that it is clearly best to prepare the state
corresponding to the largest eigenvalue of $A_{\phi_y}$ for a
given measurement outcome $y$ \footnote{The norm used in equations
(\ref{fmax2}) and (\ref{lemma}) is a special case of a $p$-norm
defined by $||A||_p=\tr{A^p}^{1/p}$. $||A||_\infty$ is just the
largest eigenvalue of $A$ and $||A||_1=\tr{A}$. For all $p'\geq p$
it holds that $||A||_{p'}\leq||A||_p$.}.

We proceed by proving a statement which is even stronger than
(\ref{bound}) namely that
\begin{equation}\label{lemma}
||A_\phi||_p\leq\frac{1+\lambda}{[(2+\lambda)^p-1]^{1/p}}||A_\phi||_1
\end{equation}
holds for all states $|\phi\rangle$ and all $p$-norms
\mbox{$||A||_p=[\tr{A^p}]^{1/p}$}. The main statement, equation
(\ref{bound}), is deduced from equations (\ref{fmax2}) and
(\ref{lemma}) by taking the limiting case
\mbox{$p\rightarrow\infty$} of equation (\ref{lemma}) in
combination with the POVM property
\mbox{$\sum_y|\phi_y\rangle\langle\phi_y|=\mathbf{1}$}, which in
turn implies $\sum_y||A_{\phi_y}||_1=1$.

In order to prove inequalities (\ref{lemma}) we exploit a trick
which was already utilized in the context of additivity of output
purities of bosonic channels in \cite{GLM}. The properties of the
trace allow us to write
\begin{eqnarray*}
||A_\phi||_p^p=\tr{A_\phi^p}&=&\int\!\!\!\int \ud\alpha_1\cdots
\ud\alpha_p p(\alpha_1)\cdots
p(\alpha_p)\\
&&\times\,|\langle\phi|\alpha_1\rangle|^2\cdots
|\langle\phi|\alpha_p\rangle|^2\\
&&\times\,\tr{|\alpha_1\rangle\langle\alpha_1|\alpha_2\rangle\cdots\langle\alpha_{p-1}|\alpha_p\rangle\langle\alpha_p|}\\
&=&\tr{|\phi\rangle\langle\phi|^{\otimes p} B},\\
||A_\phi||_1^p=\tr{A_\phi}^p&=&\tr{|\phi\rangle\langle\phi|^{\otimes
p} C}
\end{eqnarray*}
where we defined
\begin{eqnarray*}
B&=&\int\!\!\!\int \ud\alpha_1\cdots \ud\alpha_p p(\alpha_1)\cdots
p(\alpha_p)\langle\alpha_1|\alpha_2\rangle\cdots\langle\alpha_p|\alpha_1\rangle\\
&&\times\,|\alpha_1\rangle\langle\alpha_1|\otimes\cdots\otimes|\alpha_p\rangle\langle\alpha_p|,\nonumber\\
C&=&\bigotimes_{i=1}^p\int\!\ud\alpha_i
p(\alpha_i)\,|\alpha_i\rangle\langle\alpha_i|.
\end{eqnarray*}
These two operators commute evidently and thus can be diagonalized
in the same basis. The diagonalization can be accomplished
following the methods of \cite{GLM}. One finds that both $B$ and
$C$ can be expressed as a tensor product of (unnormalized) thermal
 states in certain Fourier modes attained from the original
$p$ modes by a unitary transformation. Both operators are diagonal
in the Fock basis corresponding to these new modes and are
explicitly given by
\begin{eqnarray*}
B&=&\frac{\lambda^p}{(2+\lambda)^p-1}\bigotimes_{i=1}^p\sum_{n_i=0}^\infty\left(\frac{1}{2+\lambda-d_i}\right)^{n_i}|n_i\rangle\langle
n_i|,\\
C&=&\left(\frac{\lambda}{1+\lambda}\right)^p\bigotimes_{i=1}^p\sum_{n_i=0}^\infty\left(\frac{1}{1+\lambda}\right)^{n_i}|n_i\rangle\langle
n_i|
\end{eqnarray*}
where $d_i\in\mathbb{C}$ are the eigenvalues of a unitary matrix
such that $|d_i|=1$. The exact values can easily be calculated but
are of no relevance here.

Finally, let a product state $|\phi\rangle^{\otimes p}$ have an
expansion in terms of Fock states given by
\mbox{$|\phi\rangle^{\otimes p}=\sum_{n_1,\ldots,n_p}
c_{n_1,\ldots,n_p}|n_1,\ldots,n_p\rangle$}. By construction we
know that \mbox{$0\leq\tr{|\phi\rangle\langle\phi|^{\otimes p}
B}$} and therefore
\begin{align*}
&\tr{|\phi\rangle\langle\phi|^{\otimes p} B}=\\
&=\frac{\lambda^p}{(2+\lambda)^p-1}\left|\sum_{n_1,\ldots,n_p=0}^\infty\prod_{i=1}^p\left(\frac{1}{2+\lambda-d_i}\right)^{n_i}|c_{n_1,\ldots,n_p}|^2\right|\\
&\leq\frac{\lambda^p}{(2+\lambda)^p-1}\sum_{n_1,\ldots,n_p=0}^\infty\prod_{i=1}^p\left|\frac{1}{2+\lambda-d_i}\right|^{n_i}|c_{n_1,\ldots,n_p}|^2\\
&\leq\frac{\lambda^p}{(2+\lambda)^p-1}\sum_{n_1,\ldots,n_p=0}^\infty\prod_{i=1}^p\left(\frac{1}{1+\lambda}\right)^{n_i}|c_{n_1,\ldots,n_p}|^2\\
&=\frac{(1+\lambda)^p}{(2+\lambda)^p-1}\tr{|\phi\rangle\langle\phi|^{\otimes
p} C}.
\end{align*}
Taking the $p^\mathrm{th}$ root of this sequence yields directly
relation (\ref{lemma}) and completes the proof.

The result assures that (in the case of a flat distribution) the
fidelity limit of $1/2$ is in fact appropriate in comparing
quantum channels for coherent states of continuous variables with
an optimal classical channel, justifying its application as a
benchmark in continuous variable teleportation \cite{FBFKP, ZGCLK,
BTBSRBSL, YAF} ex post and in future experiments testing the
performance of continuous variable quantum memories
\cite{KMP,SJSP, JSCFP, DP}. In particular, in a recent
experimental demonstration of the quantum state transfer from
light onto atoms \cite{JSCFP} the bound (\ref{bound}) has been
used to demonstrate that the quantum memory has indeed exceeded
the classical limit of the measure-and-prepare strategy. The
present proof provides firm grounds for such a statement.

We note that  a measure-and-prepare scheme can be considered as a
1-to-$\infty$ cloning machine, when we just duplicate the
preparation device. In fact, in this context for the special case
of a flat distribution $(\lambda\rightarrow 0)$ an independent
proof based on the covariance of the problem is given in
\cite{KWM}.

The criterion derived here allows to test wether a given channel
yields a higher quality of storage or transmission (measured in
terms of the average fidelity) than what is possible by classical
means. We would like to point out that there exist other criteria
in the literature \cite{KWM, CIR, GG, CW} allowing to test
different requirements. In particular if a channel has to be
secure (in the sense that its action excludes the existence of a
clone of the input state holding a higher fidelity than the
channel's output) it has to outperform the best 1-to-2 cloning
machine, which is more demanding than what was considered here
\cite{GG}. For channels acting on the set of coherent states with
a flat distribution this was analyzed in \cite{CIR, GG, KWM}. As
shown in \cite{CIR, GG} the best Gaussian 1-to-2 cloning machine
yields a fidelity benchmark of 2/3 while the optimal non-Gaussian
strategy yields a value of $\approx 0.6826$ as was derived in
\cite{KWM}.

Finally, we would like to point out that an experimental
demonstration of a fidelity larger than 1/2 does not disprove the
existence of a classical model in the sense of a local hidden
variable theory able to describe the physical process \cite{CW,
BFKL}. When we claim that it does prove the non-classicality or
quantumness of the respective channel then this has to be
understood in the sense that no classical measure-and-prepare
strategy can give the same result within the framework of quantum
mechanics.

In conclusion, we presented and proved a tight upper bound on the
average fidelity achievable by a classical channel for coherent
states of continuous variables subject to a Gaussian distribution
over the phase space. This limit has to be surpassed by a quantum
channel in order to outperform any competing classical strategy
and is thus of direct experimental relevance in quantum
teleportation of and quantum memories for continuous variables.
The presented result in particular validate the outstanding
experimental achievements in storing and teleporting continuous
variable quantum states. The techniques, which led to the proof of
the bound, in principle apply also to other sets of states in both
continuous variable and finite dimensional systems. Depending on
the considered sets and distributions they might thus yield
similar quantum benchmarks in other contexts.

M.M.W. thanks Reinhard F. Werner and Ole Kr\"uger for valuable
discussions. This work was supported by the EU project COVAQIAL,
 the Danish National Research Foundation and the
\mbox{Kompetenznetzwerk} Quanteninformationsverarbeitung der
Bayerischen Staatsregierung.

\end{document}